\documentclass[prb,twocolumn]{revtex4}

\usepackage{graphicx}

\begin{document}

\title{Physics of cuprates with the two-band Hubbard model - The validity of the one-band Hubbard model}
\author{A.\ Macridin} \affiliation{University of Cincinnati,
  Cincinnati, Ohio, 45221, USA} 
\author{Th.\ Maier} \affiliation{Computer Science
  and Mathematics Division, Oak Ridge National Laboratory,\ Oak Ridge,
  Tennessee 37831, USA }
\author{M. Jarrell} \affiliation{University of Cincinnati, Cincinnati,
  Ohio, 45221, USA} 
\author{G.\ A.\ Sawatzky}
\affiliation{University of British Columbia, 6224 Agricultural Road, Vancouver,  BC V6T 1Z1,
Canada}

\date{\today}

\begin{abstract} We calculate the properties of the two-band Hubbard model using the
	Dynamical Cluster Approximation. The phase diagram resembles the generic
	phase diagram of the cuprates, showing a strong asymmetry with respect to
	electron and hole doped regimes, in agreement with experiment. Asymmetric
	features are also seen in one-particle spectral functions and in the charge,
	spin and d-wave pairing susceptibility functions. We address the possible
	reduction of the two-band model to a low-energy single-band one, as it was
	suggested by Zhang and Rice. Comparing the two-band Hubbard model properties
	with the single-band Hubbard model ones, we have found similar low-energy
	physics provided that the next-nearest-neighbor hopping term $t'$ has a
	significant value($t'/t \approx 0.3$).  The parameter $t'$ is the main
	culprit for the electron-hole asymmetry.  However, a significant value of
	$t'$ cannot be provided in a strict Zhang and Rice picture where the extra
	holes added into the system bind to the existing Cu holes forming local
	singlets. We notice that by considering approximate singlet states, such as
	plaquette ones, reasonable values of $t'$, which capture qualitatively the
	physics of the two-band model can be obtained. We conclude that a
	single-band t-t'-U Hubbard model captures the basic physics of the cuprates
	concerning superconductivity, antiferromagnetism, pseudogap and
	electron-hole asymmetry, but is not suitable for a quantitative analysis or
	to describe physical properties involving energy scales larger than about
	0.5 eV.

\end{abstract}

\maketitle

\section{Introduction}
\label{sec:intr}

The theory of the cuprate high temperature superconductors remains one of the most
important and daunting problems in condensed matter physics.   The high $T_c$
cuprate superconductors are layered  materials with relatively complex structures
and  chemical composition.  They are  highly correlated, with an effective bandwidth
roughly equal to the effective local Coulomb interaction. The short-range
correlations are known to play a paramount role in these materials.  Therefore, the
Dynamical Cluster Approximation (DCA)~\cite{DCA}, which treats short-range
correlations explicitly and the long-range physics at the mean-field level, is an
ideal tool for the investigation of these systems.

A common characteristic all the cuprate materials share is the presence of
quasi-two-dimensional  $CuO_2$ planes. These planes are commonly believed to
contain the low-energy physics. However,  the full complexity of the orbital
chemistry of just the $CuO_{2}$ planes and the strong Coulomb repulsion on the Cu
ions  would lead to models which are very difficult to study with conventional
techniques. 

The cuprates are characterized by a very rich, but also in many respects  very
intriguing physics. The undoped materials are antiferromagnetic (AFM) insulators
with a gap of approximatively $2~eV$. Upon doping the AFM is destroyed and the
system becomes superconducting (SC). At small doping, in the proximity of the AFM
phase, the normal state physics cannot be described in terms of Fermi liquid theory
and is characterized by the presence of a  pseudogap.  An essential demand of every
successful theory is to capture all these fundamental features at the same time.

Experimental data show that the phase diagram and other physical
characteristics like the density of states (DOS) near the Fermi level of the
hole-doped and electron-doped materials are very
different~\cite{dagotto,hpes,armitage}.  There could be many reasons for this
asymmetry.  The electron and the hole doped materials are physically different and
apart from the $CuO_{2}$ planes, they contain different elements and chemical
structures.  These structural and compositional differences can influence the
low-energy physics.  Therefore in this paper, we use DCA to address whether the
physics of a pure $CuO_{2}$ plane contains this asymmetry or if the origin of the
asymmetry in real materials comes from other influences. 

Different models for describing the physics of a $CuO_2$ plane were proposed by
various authors.  Photoemission experiments in the insulating parent material show
that the first electron-removal states have primarily oxygen character; whereas, the
first electron-addition states have $d$ character, already suggesting a strong
asymmetry.  This places these materials in the charge-transfer gap region of the
Zaanen-Sawatzky-Allen scheme~\cite{ZSA}.  Early on, considering  the ligand field
symmetry and  band structure calculations~\cite{schuttler,mattheis,freeman}, it was
realized that the most important degrees of freedom are the Cu $d_{x^2-y^2}$ which
couple with the  in-plane O $p$ orbitals.  Therefore, one of the first models
proposed to describe the physics of high $T_c$ materials was the so called {\em
three-band Hubbard model} introduced by Varma {\em et al.}~\cite{varma3b} and Emery
{\em et al.}~\cite{emery1}, which considers explicitly both the  oxygen $p_{\sigma}$
and the  cooper $d_{x^2-y^2}$ orbitals.  In fact, because the direct oxygen-oxygen
hopping  is neglected, only the combination of oxygen orbitals with ${x^2-y^2}$
symmetry couples with the $d$ orbitals, and the above proposed three-band model
reduces to a two-band one.
 
However Zhang and Rice (ZR)~\cite{ZR} argue that the low-energy physics of the {\em
hole doped} superconductors can be described by a single-band model.  Starting from
the two-band model, Zhang and Rice claim  that an extra hole added into the oxygen
band binds strongly with a hole on the $Cu$, forming an on-site singlet.  This
singlet state, which  has zero spin can be thought as  moving through the lattice
like a hole in an antiferromagnetic background. Consequently, the physics can be
described by a one-band t-J  model.

Pertinent criticism to these simplified models were raised by various authors.  With
respect to  Cu degrees of freedom, H. Eskes {\em et al.}~\cite{expclus} stressed the
possible importance of  the other $d$ orbitals, showing that they should be
explicitly considered when physics  which implies  excitations with energy larger
than $\approx 1~eV$ is involved. However, these criticisms do not concern us for the
present study since we are interested only  in  physics at energies lower than
$\approx 0.5~eV$.

Investigating the relative importance of various parameters describing the $CuO_2$
planes it was realized early on that in addition to the Cu on-site Coulomb repulsion
($U_{dd} \approx 8 ~ eV$) and Cu-O hopping integral ($t_{pd} \approx 1.3-1.5~ eV$)
the O-O hopping integrals result in a large O $2p$ band width ($W \approx 5~ eV$),
indicating that these should be included explicitly in any
theory~\cite{expclus,stechel,Hybersten}. Therefore, using the DCA technique as a
means of including all these most important parameters and bands, we address two
major problems in this paper: The physics of the $CuO_2$ plane (including a detailed
study of the electron-hole asymmetry) and the reduction of the multi-band model to a
single-band model.

Regarding the reduction to a one-band model, one of the most serious criticisms to
ZR theory is  the neglect of the O $2p$ band structure~\cite{zaanenoles,eskesprl}.
The natural tendency of the finite oxygen bandwidth is to delocalize and destabilize
the ZR singlets. The question arises whether the low-energy states (i.e. the ZR
singlets) are still well separated from the higher energy states (i.e. the
non-bonding oxygen states). Otherwise, the reduction to a single-band model which
neglects these high energy states is not possible.  This problem  was previously
considered by Eskes and Sawatzky~\cite{eskesprl} within an impurity calculation
approach, but there, unlike in the DCA approach, both the spatial correlation
effects and the dispersion of the low-energy states were neglected. 

Another important objection to ZR theory was  first raised by Emery and
Reiter~\cite{ER}, and regards the nature of the  low-energy states.  Are these
states real singlets which can be mapped onto holes, or does the hole on the O bind
into a more complicated state which involves more than one Cu hole?  Choosing  a
particular solvable  example, which considers the Cu spins arranged
ferromagnetically, they showed that the low-energy states are in fact an  admixture
of the Zhang-Rice singlets and the corresponding triplets.  This implies a nonzero
value for the oxygen spin, and destroys the equivalence of these states to holes.
However, it is not clear if the situation is similar in the cuprates, i.e. if the ZR
singlet-ZR triplet admixture is significant. But the merit of Emery and Reiter is to
emphasize that the fact that, as a consequence of the strong Cu-O hybridization
low-energy states well separated from the non-bonding oxygen band states appear,
does not necessarily mean that the physics can be reduced to a single-band  model.

The third problem we address regarding  the reduction to a single-band model is the
estimation of the single-band parameters.  We notice that different approximations
result in different values of the parameters.  Especially the magnitude  of the
next-nearest-neighbor hopping is very dependent of the initial assumptions.  For
example, if we assume that the hole addition low-energy states are genuine ZR
singlets, i.e bound states between a Cu hole and a orthogonal Wannier  oxygen
orbital, we obtain a negligible  next-nearest-neighbor hopping~\cite{feiner}.  On
the other hand, if we consider the low-energy states to be plaquette singlets, i.e.
bound states between a Cu hole  and a hole on the  state formed by the four oxygen
around the Cu, the value of the next-nearest-neighbor hopping is
significant~\cite{eskes}.  Of course, because of the nonorthogonality of the
plaquette states, the plaquette singlets are not genuine singlets and therefore they
cannot be rigorously mapped into holes. However, because  their overlap with the
local singlets is large ($96\%$)~\cite{ZR,ER}, it is still possible that this
approximation is good. 

Our calculations show that a multi-band  model and a  single-band t-t'-U Hubbard
model  with a  significant value of the next-nearest-neighbor hopping  exhibit a
similar low-energy physics. The essential parameter needed for the agreement is the
next-nearest-neighbor  hopping, $t'$.  This parameter
is also the main culprit for the observed electron-hole asymmetry. However, as
mentioned above, the large value of $t'$ cannot be obtained in a strict ZR picture.
Thus our results also implicitly indicate that the multi-band model cannot be
rigorously reduced to a single-band model. Therefore, besides showing the
similarities between the two models, we also point out their significant differences
in this paper.

The final conclusion is that a single-band t-t'-U Hubbard model, with a significant
value of $t'$, captures the basic physics of the cuprates, and thus is suitable to
describe the AFM, pseudogap and SC physics together with the relevant asymmetries
observed in the phase diagram, in the one-particle spectra and in the two-particle
response functions. However, we believe that it  is not suitable  for a quantitative
material specific analysis, for describing the higher energy spectroscopic features
as in optical spectroscopy or resonant inelastic x-ray scattering, or for studying
more subtle features related to the finite value of the spin on the oxygen.

This paper is organized as follows. In Section~\ref{sec:form} the two-band Hubbard
model and the DCA technique is introduced.  Our two-band model takes fully into
account the oxygen dispersion and considers only the oxygen degrees of freedom which
couple directly to the Cu $d_{x^2-y^2}$ orbitals.  The results of the DCA
calculation applied to the two-band Hamiltonian are presented in
Section~\ref{sec:2bm}. The possible reduction of the two-band model to a single-band
one, together with a detailed analysis of the single-band t-t'-U Hubbard model, is
addressed  in  Section~\ref{sec:1bm}.  A discussion regarding the similarities and
the differences between the two-band model and the single-band one is given in
Section~\ref{sec:disc}.  The conclusions of our study are reviewed in
Section~\ref{sec:conc}.

\section{Formalism}
\label{sec:form}

\subsection{The model Hamiltonian}
\label{ssec:ham}

Band structure calculations~\cite{Hybersten,mcahan}, cluster
calculation~\cite{expclus}, photoemission~\cite{expclus} and other experiments show
that the relevant Cu degrees of freedom are the $d_{x^2-y^2}$ orbitals which couple
with the in-plane $p_x$ and $p_y$ O orbitals. All these degrees of freedom  result
in a five-band  (four oxygen and one copper band) Hamiltonian in general. We have
studied the five-band model in detail~\cite{macridin} and have found that due to the
strong  Cu-O hybridization, only the oxygen degrees of freedom which couple directly
with  Cu are  relevant for the low-energy physics. Consequently, to a very good
approximation, the  five-band model can be reduced to a two-band one.

The two-band model contains one Cu  $d_{x^2-y^2}$ 
correlated band and one oxygen band.  At every site the oxygen states  are obtained
by taking a linear combination with $x^2-y^2$ symmetry of the four O $p_\sigma$
orbitals which form a plaquette  around the Cu ion. These are the only oxygen states
which hybridize directly with  Cu.  However, it should be mentioned that these
plaquette states are not orthogonal, two neighboring states sharing a common oxygen
atom. An orthogonal basis can be obtained by the procedure described in the original
ZR paper~\cite{ZR}. First, applying a Fourier transform, translational invariant
(Bloch) states are constructed.  The Bloch states are orthogonal but not normalized,
so  they should be  multiplied by a normalization factor
\begin{equation}
\label{eq:bk}
\beta(k)= [\sin^2 (k_x/2)+\sin^2 (k_y/2]^{-1/2}~.
\end{equation}
\noindent After normalization a complete and orthonormal set of oxygen states 
is obtained.

In this basis the two-band Hubbard Hamiltonian can be written as:
\begin{eqnarray}
\label{eq:ham}
H=\sum_{k,\sigma} E(k) c^{\dagger}_{k\sigma}c_{k\sigma} & + & E_d  d^{\dagger}_{k\sigma}d_{k\sigma}
+ V(k) (c^{\dagger}_{k\sigma}d_{k\sigma}+h.c) \nonumber \\ 
& + & U \sum_{i}{n_{di\uparrow}n_{di\downarrow}} 
\end{eqnarray}

\noindent We  work in the hole representation and $d^{\dagger}_{k\sigma}$ ( $c^{\dagger}_{k\sigma}$)
represents the creation operator of a  Cu (O) hole with spin $\sigma$ and momentum $k$.
The O band
dispersion and the Cu-O hybridization are given by
\begin{equation}
\label{eq:epk}
E(k)=E_{p}-8 t_{pp} \beta^2(k) \sin^2 (k_x/2) \sin^2 (k_y/2)
\end{equation}
\begin{equation}
\label{eq:vk}
V(k)=2 t_{pd} \beta^{-1}(k)
\end{equation}

\noindent with $t_{pp}$ being the O-O hopping integral.  The last term in
Eq.\ref{eq:ham} represents the Coulomb repulsion between two holes on the same $d$
orbital. We choose the commonly accepted values of the parameters, based on the band
structure calculations of McMahan {\em et al.} ~\cite{mcahan} and Hybertsen {\em et
al.}~\cite{Hybersten} Because of the low density of oxygen holes ($25\%-30\%$), we
treat the Coulomb repulsion on $O$ (given by $U_{pp}$) and the repulsion between
nearest-neighbor Cu and O holes (given by $U_{pd}$) at the mean field level as a
reasonable approximation.  The effect will be an increases of our estimation for
$\Delta=E_p-E_d$ by $U_p \frac{\overline{n}_{p}}{2}+
U_{pd}(\overline{n}_{d}-\overline{n}_{p})$, where $\overline{n}_{d}$ and
$\overline{n}_{p}$ are the average occupation of Cu and respectively O bands. A
choice of $U_{pp}=6~eV$, $U_{pd}=1.3~eV$ and $\overline{n}_{p}=0.3$ results in a
increase of $\Delta$ by $1.3~eV$.

To conclude, we take in
Eq.~\ref{eq:ham},  $t_{pd}=1.3~eV, ~t_{pp}= 0.65~eV, ~\Delta=4.8~ eV$ and $U=8.8~ eV$.

\subsection{DCA technique}
\label{ssec:dca}

\begin{figure}[!t]
\centerline{
\includegraphics*[width=2.5in]{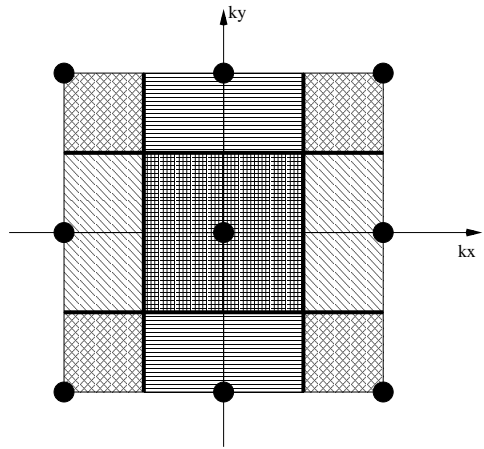}}
\caption{ Coarse-graining of the Brillouin Zone in four cells ($N_c=4$) around 
 $K$=$(0,0),(0,\pi),(\pi,0)$ and $(\pi,\pi)$.}
\label{fig:coarse}
\end{figure}

The DCA is an extension of the Dynamical Mean Field Theory (DMFT)\cite{DMFT}.  The
DMFT maps the lattice problem to an impurity embedded self-consistently in a host
and therefore neglects spatial correlations. The DCA maps the lattice to a
finite-sized cluster embedded in a host.  Non-local correlations up to the cluster
size are treated explicitly, while the physics on longer length scales is treated at
the mean field level. This
technique is ideal for the problems where short-ranged correlations are predominant
as in the high-$T_c$ materials.  Here we calculate the properties of the embedded cluster with
a Quantum Monte Carlo (QMC) algorithm. The cluster self-energy is used to calculate
the properties of the host, and this procedure is repeated until a self-consistent
convergent solution is reached.  The self-energy and vertex functions of the cluster
are then used to calculate lattice quantities.   A
detailed description of the QMC-DCA algorithm is given in ref.~\cite{jarrelldca}

In the two-band model the oxygen degrees of freedom are not correlated, and
therefore they are not included explicitly in  the cluster. Their effect is fully
contained in the cluster-host hybridization function and in the host Green's
function.

Here we consider a $2 \times 2$ cluster of Cu ions, which we believe to be large
enough to capture the essential physics of Hubbard-type models.  The $2 \times 2$
cluster will result in a coarse-graining of the BZ in four cells, as shown in
Fig.~\ref{fig:coarse}.

\section{Two-band Hubbard model results}
\label{sec:2bm}

\begin{figure}
\includegraphics*[width=3.3in,height=2.2in]{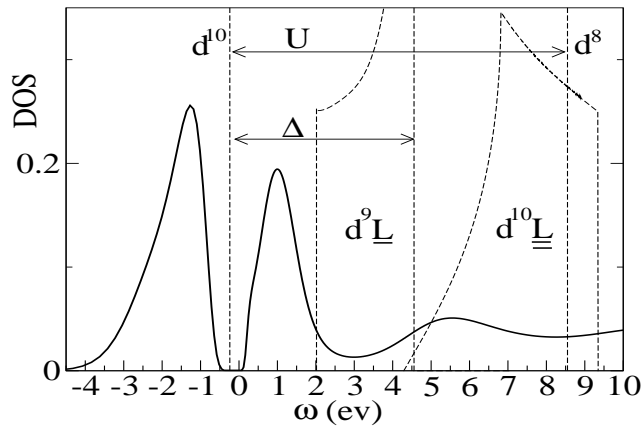}
\caption{ Two-band Hubbard model DOS at $0\%$ doping. 
The solid line is the $d$ part of the DOS 
calculated at $T=685~ K$.
The value of the  parameters  is 
$t_{pd}=1.3eV, ~t_{pp}=0.65eV, ~\Delta=4.8 eV,~ U=8.8 eV$.
The dashed line shows the  DOS when $t_{pd}=0$. The chemical potential
$\mu=0$.}
\label{fig:zr}
\end{figure}

The undoped materials have one hole per $CuO$ unit.  For $t_{pd}=0$ the DOS is given
by the dashed line in Fig.~\ref{fig:zr} and the hole addition states would be of
pure O character.  When the Cu-O hybridization $t_{pd}$ is switched on, the extra
holes added to the oxygen band will scatter with the $Cu$ spins and bound states
will appear at the bottom of the oxygen band.  This is illustrated by the solid
line, which plots the partial $d$ DOS which was obtained using the Maximum Entropy
Method (MEM)\cite{MEM} for the analytic continuation of the QMC data to real
frequencies.  It can be noticed that the first hole addition states have a strongly
mixed $d$ and $p$ character (the $d$ character in the spectrum is large now) and an
energy pushed well below the edge of the initial non-bonding oxygen band.  Therefore
only these states are relevant for the low-energy physics~\footnote{ A five-band
Hubbard model calculation~\cite{macridin} confirms that, the occupation number of
the non-bonding oxygen bands is less than $1\%$ up to $40\%$ hole doping.}. In the
ZR theory these low-energy states which appear as a consequence of the strong Cu-O
hybridization are considered to be local singlets which move through the lattice
like holes in an AFM background. Consequently the claim is that the physics can be
described by a one-band t-J model.

\begin{figure}[t]
\includegraphics*[width=3.3in,height=2.2in]{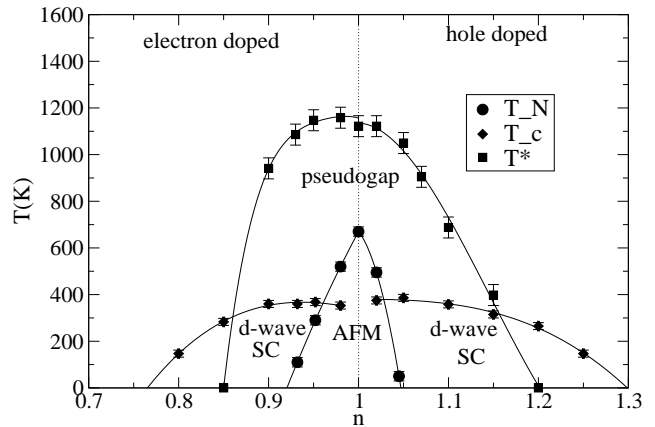}
\caption{ Two-band Hubbard model phase diagram.}
\label{fig:phase}
\end{figure}

In order to determine the phase diagram we calculate a large number of
susceptibilities which are relevant for spin, charge and superconducting ordering,
both at the center and the corner of the BZ. For example, the N\'eel and SC critical
temperatures, $T_{\rm N}$ and respectively $T_c$ in the phase diagram presented in
Fig.~\ref{fig:phase}  are determined from the divergence of the corresponding
susceptibilities. The pseudogap crossover temperature $T^*$ is obtained from the
maximum in the uniform magnetic susceptibility when accompanied by a suppression of
spectral weight in the DOS.  Similar to what was found in the single-band Hubbard
model~\cite{jarrellhub}, we find AFM and d-wave SC for both electron and hole doped
regimes.  However the electron-hole symmetry is broken.  In the electron doped case
AFM persists to a much larger doping.  On the contrary, SC disappears at a smaller
critical doping~\footnote{ Our preliminary calculations indicate that the
electron-hole asymmetry in the phase diagram is  more pronounced when  larger than
$2 \times 2$ clusters are considered.}. These features of the phase diagram are in
qualitative agreement with the experimental findings~\cite{dagotto}.

\begin{figure}
\includegraphics*[width=3.3in,height=2.5in]{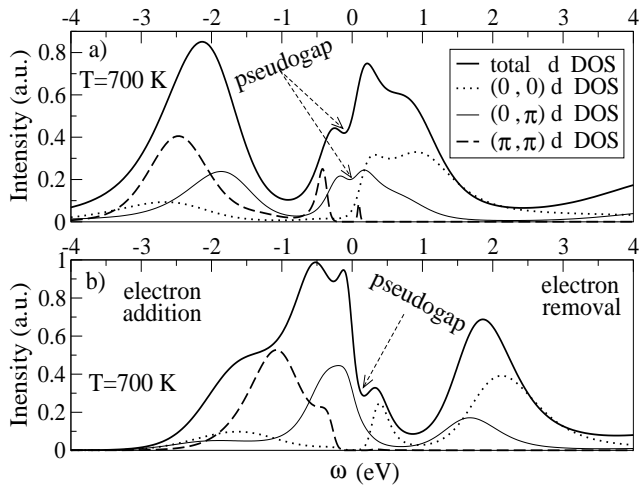}
\caption{Total $d$ DOS and coarse-grained $K$ dependent $d$ DOS at $5\%$ doping.
a)  hole doping case. b)  electron doping case.}
\label{fig:dos}
\end{figure}

The one-electron spectral functions as measured with photoemission are also
different.  Our $2\times 2$ cluster divides the BZ into four cells around
$K$=$(0,0),(0,\pi),(\pi,0)$ and $(\pi,\pi)$ (see  Fig.~\ref{fig:coarse}) and
approximates the lattice self-energy by a constant $\Sigma(K,\omega)$ within a cell.
Because of this coarse-graining, a comparison with ARPES is not possible apart from
gross features.  However, as the phase diagram shows, we believe that even our small
cluster captures much of the physics of the cuprates.  Here we want to stress the
difference between the electron and the hole doped case within our $2\times 2$
cluster approximation.  In Fig.~\ref{fig:dos} a) and b) we show the total $d$ states
DOS and the $d$ coarse-grained $K$ dependent DOS (which would correspond to the
average over all $k$ belonging to a coarse-graining cell of the single particle
spectra $A(k,\omega)$) for the hole and respectively for the electron doped case, at
$5\%$ doping. The total DOS looks qualitatively similar and at the chemical
potential we see in both cases a depletion of states which indicates the presence of
the pseudogap.
The $K$ dependent DOS is very different.  The important feature which we want to
stress is the location of the pseudogap in the BZ.  In the hole doped case the
pseudogap appears around $(0,\pi)$.  For the electron doped case we do not detect
any suppression of states around $(0,\pi)$ even though the pseudogap is clearly
present in the total DOS.  These features are in agreement with the photoemission
experiments.  The hole doped materials show Fermi pockets around $(\pi/2,\pi/2)$ and
gapped states around $(0,\pi)$~\cite{hpes}. For the electron doped materials the
photoemission spectra~\cite{armitage} exhibit a gap near $(\pi/2,\pi/2)$ and Fermi
surface pockets around $(0,\pi)$.  With the present cluster size the DCA cannot
determine where in $k$-space the pseudogap is, but it is interesting that it is not
at  $(0,\pi)$. The presence of the pseudogap at $(\pi/2,\pi/2)$ for the electron
doped system can only be checked by increasing the cluster size, and this work is in
progress.

\begin{figure}[t]
\includegraphics*[width=3.3in,height=2.5in]{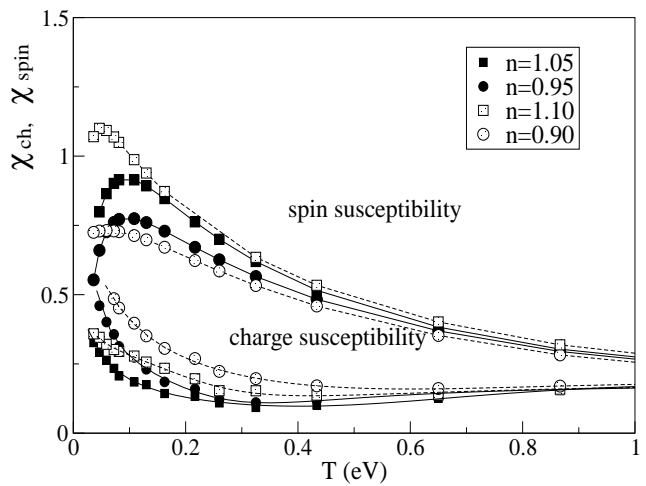}
\caption{Uniform spin $\chi_{spin}$  (upper part) and charge $\chi_{ch}$ (lower part)  susceptibilities versus 
temperature for different hole densities. $n$ in the legend represents the number of holes per unit cell.}
\label{fig:sus}
\end{figure}

\begin{figure}
\centerline{
\includegraphics*[width=3.3in,height=2.5in]{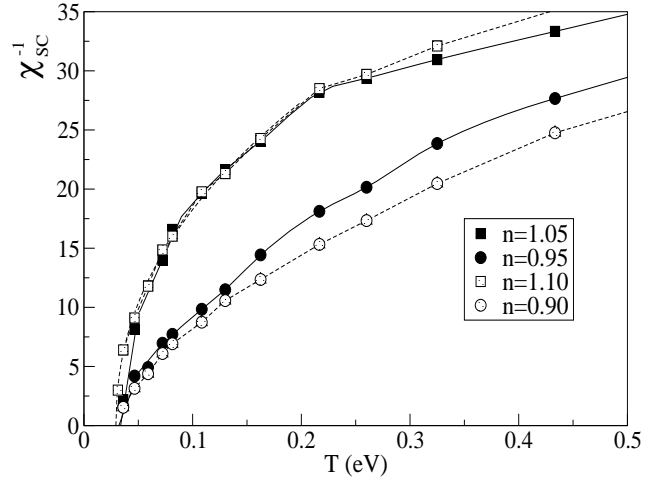}}
\caption{Inverse of the $d$-wave pairing  susceptibility $\chi^{-1}_{SC}$ versus temperature
for different hole densities.}
\label{fig:scsus}
\end{figure}   

\begin{figure}
\centerline{
\includegraphics*[width=3.3in,height=2.5in]{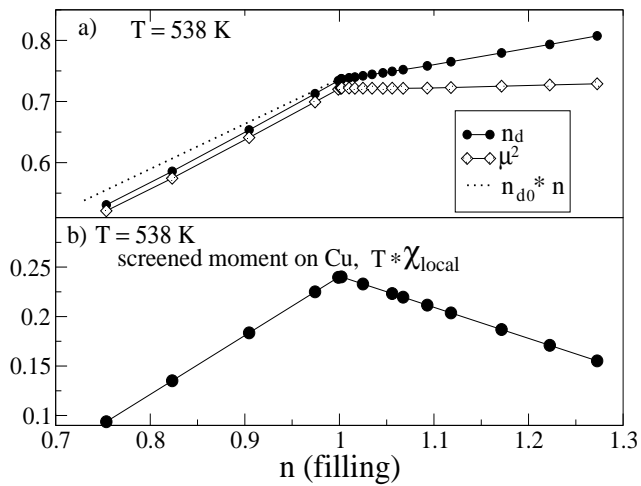}}
\caption{a) The Cu occupation number, $n_d$, the unscreened Cu moment,
 $\mu^2$ (Eq.~\ref{eq:uscr})  versus hole filling. b) The 
screened Cu moment, $T~\chi_{local}$ (Eq.~\ref{eq:scr}) versus hole filling.}
\label{fig:ndmiu}
\end{figure}

The electron and the hole doped susceptibility functions are also different both for
the divergence temperatures and the temperature and doping dependence.  In
Fig.~\ref{fig:sus}  we show the uniform spin and charge susceptibilities versus
temperature at $5\%$ and $10\%$ doping.  A common feature for all cases is the
existence of a characteristic temperature $T^{*}$ below which the spin response is
suppressed and the charge one is enhanced.  $T^{*}$ corresponds to the pseudogap
(seen in the DOS) onset temperature.  The suppression of the spin excitations below
$T^{*}$ was also seen in NMR~\cite{nmr} experiments and it was associated with the
pseudogap.  Besides these common features the electron and the hole doped
susceptibilities behave differently. Generally, the maximum value of the spin
susceptibility increases with hole filling. This means that in the hole doped case,
the spin susceptibility at the pseudogap temperature is strongly increasing with
doping unlike in the electron doped case where it decreases upon doping.  At the
same doping the hole doped spin susceptibility is much larger than the electron
doped one.  Another interesting feature is the very strong increase of the charge
susceptibility for the electron doped case in the underdoped region ($5\%$ doping),
suggesting a tendency towards phase separation~\cite{macridin2}. 

Asymmetric behavior can also be noticed in Fig.~\ref{fig:scsus}, where we  plot the
{\em inverse}  of the  $d$-wave paring susceptibility.  Above $T_{c}$ the pairing
susceptibility increases with doping in the electron doped case and remains more or
less constant in the hole doped case.

Because of the large Cu-O hybridization the system is strongly covalent.  For
example in the undoped regime the Cu occupation number is only $\approx 73\%$.   The
fact that the cuprates are strongly covalent was also observed in NMR
measurements~\cite{nmreskes}.  We  notice that the system exhibits a slightly doping
dependent covalency.  This is shown in  Fig.~\ref{fig:ndmiu}-a where the Cu
occupation number versus hole density is plotted.  A constant covalency, equal to
the one in the undoped regime (i.e. $0.73$ Cu holes and $0.27$ O holes per site),
would correspond to the dashed line. It can be noticed that, for the electron doped
regime, the Cu hole occupation number is decreasing faster than the hole
concentration, which indicates an increasing covalency with increasing electron
doping.  This happens because at large electron doping, i.e. when the  hole-filling
of the $CuO_2$ plane is small, the effective hybridization is  a result of a large
$V(k)$ in the BZ ~\footnote{The  Cu-O hybridization (Eq.~\ref{eq:vk}) is strongly
$k$ dependent, its value taking values from $2\sqrt{2}t_{pd}$ at $(\pi,\pi)$ point
to zero at $(0,0)$ point in the BZ.}.  Increasing the number of holes, the BZ starts
to fill up and a smaller  $V(k)$ will be responsible for the hybridization, and
consequently the covalency  decreases.  For the hole doped regime, the extra holes
go primarily on the oxygen band, and therefore we do not have a direct measure of the
covalency. 

In Fig.~\ref{fig:ndmiu}-a the unscreened moment on the Cu orbitals is shown.  It is
defined as:
\begin{equation}
\label{eq:uscr}
 \mu^2=\langle (n_{di\uparrow}-n_{di\downarrow})^2 \rangle = 
n_d -2 \langle n_{di\uparrow}n_{di\downarrow}  \rangle~.
\end{equation}
\noindent The difference between $n_d$ and $\mu^2 $ is a measure of the
double occupancy with holes on Cu sites.  In the electron doped regime the double
occupancy is very small, but it increases substantially in the hole doped regime,
which indicates that the low-energy hole addition states contain double occupied Cu
configurations in a significant measure. 

In  Fig.~\ref{fig:ndmiu}-b the screened moment on Cu, defined as
\begin{equation}
\label{eq:scr}
 T~\chi_{local}=\frac{T}{N}\sum_{i}\int_0^\beta{\langle S^{-}_i(\tau)S^{+}_i(0)\rangle d\tau}~,
\end{equation}
\noindent where $S_i$ is Cu spin operator at site $i$, is shown.  The main effect of
the extra holes is to screen the spins on the Cu sites. The screening starts to be
effective bellow temperatures of about $\approx 0.5~eV$ (not shown).  In the
Zhang-Rice scenario an extra hole  perfectly screens one spin on Cu forming a
strongly bound on-site singlet which would contain a significant amount of the
double occupied Cu configuration.  So, our results do not contradict the ZR theory,
but also do not exclude other scenarios where the extra holes form more complicated
bound states which involve more than one Cu spin. Quantitative analysis based on the
amount of screening as function of hole doping cannot give an answer to the validity
of the ZR assumption, because, aside from the screening due to the oxygen holes,
there are also non-local processes which contribute to the screening of Cu moments
(for example, a possibility is the formation of inter-site spin singlets associated
with the Resonance Valence Bond scenario).

\section{Reduction to single-band Hubbard model}
\label{sec:1bm}

Concluding that the electron-hole asymmetry is an intrinsic property of the $CuO_2$
plane, we next address the cause of this asymmetry and the possible reduction to a
one-band model.

In Sec.~\ref{sec:2bm} we showed that, due to the Cu-O hybridization, the addition of
holes results in the formation of low-energy states, with an energy well bellow
($\approx 1~eV$) the initial oxygen band (see Fig.~\ref{fig:zr}).  The reduction to
a one-band model is based on the ZR claim that these states are singlets, i.e.
spinless entities which can be regarded as holes moving in an antiferromagnetic
background.  Due to the Monte Carlo nature of our calculation, which does not
provide a wave function for the ground state, we cannot determine directly the exact
nature of these states. The most we can do is to compare the results of a two-band
Hubbard model calculation with those of a one-band Hubbard model, and based on the
similarities and differences that we might find to decide about the validity of the
single-band approach.

\subsection{Zhang \& Rice approximation and the derivation
of the effective parameters}
\label{ssec:reduction}

In order to compare the two-band and the one-band models,
we should first get an idea about the possible single-band model effective
parameters.  We  discuss  here  two different approaches for calculating these parameters,
both based on the assumption that the low-energy states are localized,
and close to the ZR proposed singlets.

\subsubsection {Cell-perturbation theory} 
\label{sssec:cell}

\begin{table*}
\begin{center}
\begin{tabular}{|l|l|l|l|l|l|l|l|l|l|} \hline
cell-perturbation   &$U=3.04$ & $J=0.25$  & $J'\approx 0 $ &  $t^h=0.477 $ & $t^e=-0.35 $ & $t^J=0.433 $ 
& ${t^h} ' =-0.03 $ & ${t^e}' =-0.016 $ & ${t^J} '=-0.003 $ \\ \hline
cluster calculation &       & $J=0.192$ & $J'=0.012$     
&  $t^h=0.452 $ & $t^e=-0.323$ &      & ${t^h} '=-0.169$ & ${t^e}' =0.078$  &   \\ \hline
\end{tabular}
\end{center}
\caption{\label{tab:zrparam}First row: parameters in eV calculated using 
cell-perturbation theory. Second row: parameters in eV calculated using cluster
diagonalization.}
\end{table*}

The cell-perturbation theory~\cite{feiner} assumes that the ZR mapping is {\em
strictly} true and therefore the low-energy states are genuine local singlets. Here
and everywhere in the paper by local we refer to the oxygen {\em orthogonal} Wannier
states which are different from the non-orthogonal plaquette states around the Cu
ions.   

To deduce the one-band model parameters we work in the site representation.  We can
Fourier transform  Eq.~\ref{eq:ham} and write it as
\begin{equation}
\label{eq:ham0}
H=H_0+H_1
\end{equation}
\noindent where 
\begin{eqnarray}
\label{eq:ham1}
H_0=\sum_iH_{0i}=&\sum_i \sum_\sigma  (E_0 c^{\dagger}_{i\sigma}c_{i\sigma}  +  
E_d  d^{\dagger}_{i\sigma}d_{i\sigma}\nonumber \\
&+ V_0 (c^{\dagger}_{i\sigma}d_{i\sigma}+h.c)) 
 + U n_{di\uparrow}n_{di\downarrow}~.
\end{eqnarray}
\noindent Here $i$ represents the site index. The oxygen operators
$c_i$ describe the orthogonal Wannier states.
 The ZR assumption implies
that $H_0$ is responsible for the formation of the low-energy states (local singlets),
and  $H_1$ will determine the hopping parameters. 
Therefore the cell-perturbation theory provides a means to determine
the one-band parameters provided that the ZR assumption is correct. 
Elaborate calculations along this line were  done in~\cite{feiner} for a variety of
multi-band parameters.  
In a first order approximation in $H_{\rm 1}$, the effective U is given by
\begin{equation}
\label{eq:ueff}
U_{eff}=E^2+E^0-2 E^1
\end{equation} 
\noindent where $E^2$, $E^1$ and $E^0$ represent
the energies of the two (i.e. the ZR singlet), one (i.e. the bonding state) and
respectively zero hole states of Eq.~\ref{eq:ham1}.  An important point is that
$H_1$ introduce three types of hoppings. If we denote with $|2_i>$, $|1_i>$ and
$|0_i>$ the lowest energy states of $H_{0i}$ corresponding to two, one  and
respectively zero holes, we have the following hopping integrals
\begin{eqnarray} 
\label{eq:hopph}
t^h_{ij}=&<2_i,1_j|H_1|1_i,2_j> \\
\label{eq:hoppe}
t^e_{ij}=&<0_i,1_j|H_1|1_i,0_j> \\
\label{eq:hoppj}
t^J_{ij}=&<1_i,1_j|H_1|0_i,2_j>
\end{eqnarray}
$t^h$ (Eq.~\ref{eq:hopph}) describes the  hopping of the ZR singlet, $t^e$
(Eq.~\ref{eq:hoppe}) the hopping of the electron, and $t^J$ produces the exchange
interaction 
 \begin{equation}
\label{eq:jj}
J=4 {t^{J}}^2/U_{eff}~.
\end{equation}
The cell-perturbation theory applied to our model gives the parameters shown 
in the first row of Table.~\ref{tab:zrparam}.

We want to point out two things. First, the reduced Hamiltonian in the
cell-perturbation theory is a t-t'-J model,
\begin{equation}
\label{eq:ttj}
H=-t \sum_{<i,j>}\hat{b}^{\dagger}_i\hat{b}_j-t' \sum_{<<i,j>>}\hat{b}^{\dagger}_i \hat{b}_j +
J \sum_{<i,j>}S_i S_j~,
\end{equation}
with different hopping parameters for the electron and the hole doped regions and
with a value of the exchange interaction not determined by the quasiparticle's
hopping ($t^h$ or $t^e$), but, as it is shown in Eq.~\ref{eq:jj} by $t^J$.
Therefore, a comparison with a one-band Hubbard model, should be done cautiously.
Second, we want to stress that  the value of the next-nearest-neighbor hopping terms
($t'_e$ and $t'_h$) is  very small compared to the nearest-neighbor terms. The
reason is that the initial oxygen-oxygen hybridization, $t_{pp}$, results in an
effective hopping term comparable in magnitude with the one resulting from the
copper-oxygen hybridization, but with a different sign. 
This  was also remarked in~\cite{feiner}, and turns out to be an important
observation for our final conclusions.

\subsubsection{Cluster calculation} 
\label{sssec:clust}

The other approach  used for determining the parameters of the one-band model is
based on a cluster calculation. In order to estimate the nearest-neighbor hopping,
the next-nearest-neighbor hopping and the exchange terms, Eskes {\em et al.}
\cite{eskes}considered two clusters, $CuO_7$ (which contains two nearest-neighbor Cu
ions) and respectively $CuO_8$ (which contains two next-nearest-neighbor Cu ions).
The exchange term is determined as the energy difference between the singlet and the
triplet state of two holes on a cluster. For three holes on a cluster, the two
energetically lowest states can be very well ($98\%$) approximated with the bonding
and anti-bonding  states of a {\em plaquette ZR singlet} hopping between the two
cells. Therefore, the differences between these two levels is two times the ZR
singlet hopping, $t^h$. In an analogous way, considering only one hole on a cluster,
the  electron hopping, $t^e$, is determined.  Using the cluster approach,  our
two-band model results in the effective parameters shown in the second row of the
Table~\ref{tab:zrparam}. 

\subsubsection{Comparison of the two approaches}
\label{sssec:comp}

It can be immediately noticed that the two approaches produce different parameters,
especially regarding the value of the next-nearest-neighbor hoppings.  In the
cluster calculation we obtain significant  next-nearest-neighbor hoping terms,
$|{t^e}'|/|t^e|=0.22$ and $|{t^h}'|/|t^h|=0.37$ with different sign for the hole and
respectively electron doped case (${t^h}'<0$, ${t^e}'>0$).

The reason for the discrepancy between the two approaches is that, unlike the
cell-perturbation method which considers local singlets,  the cluster approach
considers singlets  between a Cu hole and a oxygen state formed on the plaquette
around the Cu ion.  Since the oxygen plaquette states are nonorthogonal it is
possible to write them as a linear combination of many orthogonal oxygen states at
different sites, i.e. the plaquette singlets are nonlocal states (in the orthogonal
base).  It should be pointed out that  at  first glance this non-locality seems
irrelevant (the overlap of the local oxygen states with the plaquette
states~\cite{ZR,ER} is about $96\%$), but apparently it turns out to influence
the value of the next-nearest-neighbor hopping of the reduced
Hamiltonian considerably.

As pointed out by Emery and Reiter~\cite{ER}, the plaquette singlets are in fact an
admixture of ZR singlets and ZR triplets and this can result in a significant value
of the spin-spin correlation on the oxygen sites~\footnote{In their example, Emery
and Reiter considered a ferromagnetic alignment of Cu spins and showed that the  ZR
singlet -ZR triplet admixture is even at the (0,0) point in the Brillouin Zone.}.
Of course, if the ZR singlet-ZR triplet admixture is significant, this will make a
rigorous reduction to a one-band model impossible, because now the low-energy states
have also a spin component and  therefore cannot be mapped into holes (spinless
entities).

It is worth pointing out that, in the cluster approach, the large value of the
next-nearest-neighbor hopping terms results solely from the finite oxygen dispersion
and the lack of hopping between the copper and the next-nearest neighbor oxygen
plaquette state. On the other hand, in the cell-perturbation theory a copper -
next-nearest neighbor oxygen hopping term is present.  It results in an effective
next-nearest-neighbor hopping with a  sign different than the one produced via
oxygen-oxygen hopping.

\subsection{Possible reasons for the reduction to fail}
\label{ssec:disc}

We believe that a comparison between the two-band Hubbard model and a single-band
Hubbard model should be done with  extreme caution. We want to stress the possible
problems here.

First, the reduction based on the ZR approximation, which results in a single-band t-J
(or t-t'-J) model assumes the strong-coupling limit, i.e. a ratio  $U_{eff}/t \gg 8$
(the two-dimensional  bandwidth is $W=8t$).  The low-energy density of states of the
two-band model shown in Fig.~\ref{fig:zr} and Fig.~\ref{fig:dos}  indicates a
bandwidth of the order of the gap, showing that we are rather at the
intermediate-coupling than at strong-coupling.  In the cell-perturbation theory we
get  $U_{eff}/t^J =7.02$, which also suggests intermediate-coupling physics.
Therefore, the question to be asked is whether the intermediate coupling
regime, characterized by an effective repulsion of the same order of magnitude as
the bandwidth, can still be well approximated by a second-order perturbation reduced
t-J model.

Second, considering  the previous objection, one may think that a reduction to the
single-band Hubbard model in intermediate coupling regime, rather than to a t-J
model, is more appropriate.  However,  serious problems arise from the fact that, in
the ZR theory the nature of the antiferromagnetic correlations is different from
that in the single-band Hubbard model, i.e. it is not directly related with the
quasiparticle (ZR singlet or electron) hopping. Therefore, unless both the two-band
Hubbard model and the one-band Hubbard model can be reduced to a t-J model, a
comparison between them does not make much sense.  Nevertheless, we believe that
even when the effective repulsion is comparable with  the bandwidth the second order
perturbation theory which produce the t-J model can be used successfully.  We are
going to discuss this at the beginning of  the next section (Sec.~\ref{ssec:ttU}).

Third, the non-locality of the low-energy states (in the sense discussed in
Sec.~\ref{sssec:comp}) can have very serious consequences beyond determining the
value of the hopping parameters, making the single-band approach to fail completely.

\subsection{t-t'-U Hubbard model results}
\label{ssec:ttU}

\begin{figure}
\includegraphics*[width=3.3in,height=2.2in]{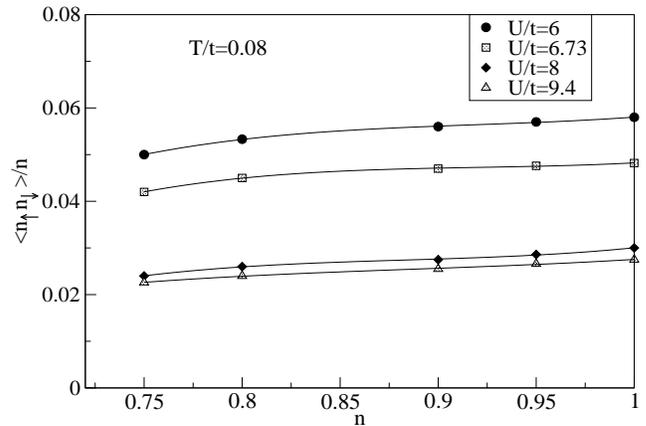}
\caption{ The relative double occupancy of the orbitals, \mbox{$<n_{\uparrow}n_{\downarrow}>/n$},
versus hole filling, $n$, for different
values of the ratio $U/t$ of the single-band Hubbard model. }
\label{fig:docc}
\end{figure}

The t-J model results as a low-energy effective Hamiltonian from the Hubbard model
by projecting out the doubly occupied states. Therefore, the  double occupancy of
the site orbitals constitutes a measure of the validity of this approximation.  In
Fig.~\ref{fig:docc} we plot the double occupancy of the site orbitals for different
values of the ratio $U/t$.  It can be noticed that for $U/t \ge 6$  the double
occupancy is always smaller than $6\%$~\footnote{ We also find (not shown) that with
decreasing  $U/t$ beyond this value, the double occupancy increases fast, being
about $15\%$ for $U/t=4$.}. This indicates that,  even in the intermediate coupling
regime the low-energy physics of the one-band Hubbard model  can be  well described
by a t-J model.

Even if, it is more natural to compare the two-band model with a t-t'-J (or a
t-t'-J-J') model, this turns out from our perspective to be rather inconvenient due
to the technical difficulties encountered by the QMC when applied to such models.
Therefore, we proceed by comparing the two-band model with a t-t'-U Hubbard model,
focusing on the qualitative features rather than on a quantitative comparison.  In
the strong-coupling limit, the t-t'-U model reduces to a t-t'-J-J' model, 
with the constraint $J'=J \times (t'/t)^2$.  Therefore, it is reasonable to assume
that if the value of $(t'/t)^2$ is not too large and the reduction of the two-band
model to a single-band model is valid, the two models should exhibit similar physics.

Assuming that the reduction to a one-band model in the spirit of the ZR approximation
is possible, we should expect from Table.~\ref{tab:zrparam} the hopping
parameters to be different in the hole doped and in the electron doped regions.
On the other hand the exchange interaction, 
\begin{equation}
\label{eq:jju}
J=\frac{4t^2}{U}~,
\end{equation}
\noindent should be  the same.

Therefore, we study  the single-band t-t'-U Hubbard model and address the following
questions: 
\begin{itemize}
\item[1)] How do the system properties depend on the ratio $t/J$?
\item[2)] What is the role of the next-nearest-neighbor hoping t'?
\end{itemize}

\subsubsection{The  t/J dependence} 
\label{sssec:tj}

\begin{figure}[t]
\centerline{
\includegraphics*[width=3.3in,height=2.2in]{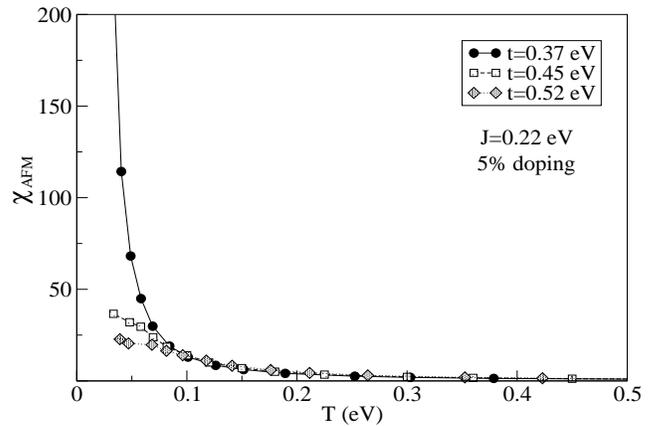}}
\caption{Antiferromagnetic susceptibility at $5 \%$ doping, for
three different values of $t$, when $J$ is constant.}
\label{fig:jtafm}
\end{figure}

\begin{figure}
\centerline{
\includegraphics*[width=3.3in,height=2.2in]{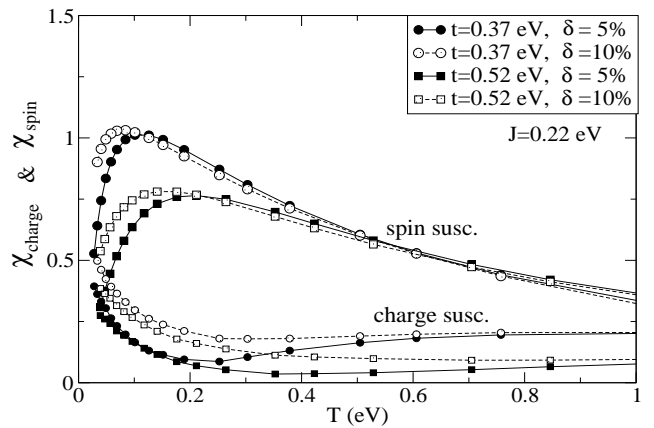}}
\caption{Spin and charge susceptibilities  at  $5 \%$ (black)
and  $10 \%$ (grey) doping for $t=0.37~eV$ (circle) and  $t=0.52~eV$ (square).}
\label{fig:jtssus}
\end{figure}

\begin{figure}[t]
\centerline{
\includegraphics*[width=3.3in,height=2.2in]{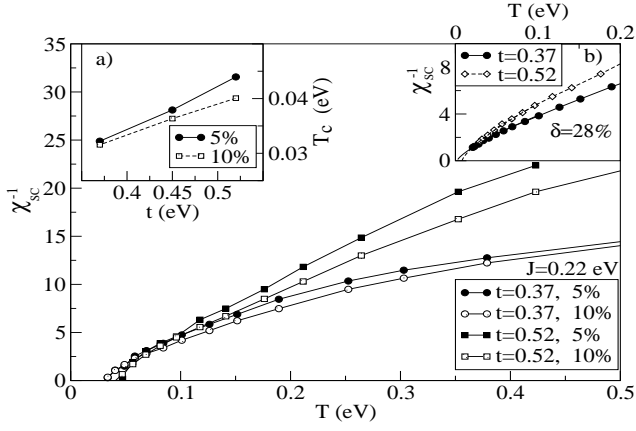}}
\caption{Inverse of d-wave pairing susceptibility versus temperature
for different hole densities and hopping parameters. Inset -a) The
critical temperature versus $t$ at $5\%$ (circle) and $10\%$ (squares) doping.
Inset -b) Inverse of d-wave pairing susceptibility versus temperature at $28\%$
doping, for $t=0.37~eV$ (circles) and $t=0.52~eV$ (diamonds). } 
\label{fig:jtdsus}
\end{figure}

\begin{figure}
\centerline{
\includegraphics*[width=3.3in,height=2in]{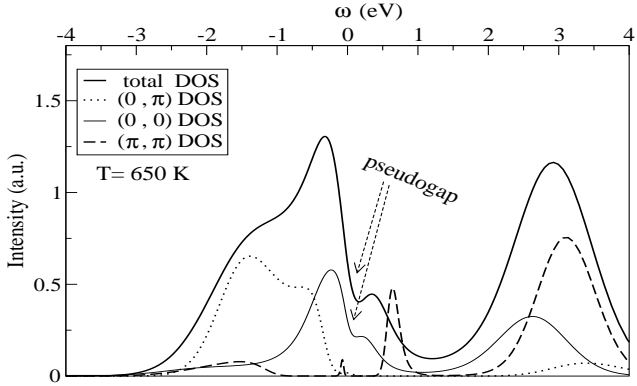}}
\caption{Single-band t-U Hubbard model total and $K$ dependent DOS at $5\%$ doping.
$J=0.22~ eV$, $t=0.45~eV$  } 
\label{fig:jtdos}
\end{figure}

The values of the parameters in Table.~\ref{tab:zrparam} show that in general the
ratio $|t/J|$ is larger in the hole-doped regime than in the electron-doped case. In
order to address the  electron-hole asymmetry observed in the two-band model, in
this section we study the properties of the single-band Hubbard model as a function of
$t/J$, by keeping $J$ (given by Eq.~\ref{eq:jju}) constant and varying the hopping
$t$. The next-nearest-neighbor hopping $t'$ is set to zero.

With respect to antiferromagnetism,  with increasing $t$ the N\'eel temperature at small doping  
and the critical doping where the antiferromagnetism disappear decrease. For example,
at $5\%$ doping, the antiferromagnetic susceptibility is diverging only for the small
value of $t$ shown in  Fig.~\ref{fig:jtafm}. 
Assuming that the hole
doped regime is characterized by a larger value of $t/J$, this feature is in
agreement with the two-band model asymmetric behavior (see Fig.~\ref{fig:phase}).

The uniform spin susceptibility is shown in Fig.~\ref{fig:jtssus}.  One can notice
that an increase of $t$ results in an increase of $T^*$ and a decrease of the spin
susceptibility at $T^*$.  This together with the behavior of the susceptibility as a
function of doping is in contrast to what was observed in the two-band model
(see Fig.~\ref{fig:sus}) where the spin susceptibility is larger in the hole doped
case and an increase (decrease) with doping of the susceptibility at $T^*$ for the
hole (electron) doped regimes is found.

The behavior of the $d$-wave pairing susceptibility as a function of $t$ is shown in
Fig.~\ref{fig:jtdsus}.  The critical temperature increases  with increasing $t$ (the
increase of $T_c$ is about $10\%$ of the increase of $t$), as can be seen in the
inset (a).  This increase is much too large to be in agreement with the two-band
model results even if, actually, for the two-band model we obtained a hole-doped
$T_c$  larger with about $20~K$  than the electron-doped one~\footnote{Nevertheless
we should take the necessary precautions saying that this value is of the order of
the error bar  (see Fig.~\ref{fig:phase}).}.  By extrapolating the inverse of the
$d$-wave pairing susceptibility at $28\%$ doping (see inset (b)) it can be concluded
that an increase of $t$ results in an increase of the critical doping where  SC
disappears. We also notice that, at small doping and above $T_c$, a large $t$ 
suppresses the pairing correlations.
These features  are in agreement with the asymmetry of 
the two-band model phase diagram.  
Nevertheless, we notice that, above $T_c$ and for both values of $t$, with increasing the doping
the pairing correlations increase too.  This behavior is characteristic in the
electron doped regime of the two-band model, but cannot explain 
the  hole doped regime where the pairing does not depend  on the doping 
(see Fig.~\ref{fig:scsus}).
The other difference between the two-band and
the single-band Hubbard model is the value of the SC susceptibility critical exponent
$\gamma$, which is much smaller in the two-band model case. 

The density of states and the $K$ dependent DOS for the one-band t-U Hubbard model
at $5\%$ doping is shown in Fig.~\ref{fig:jtdos}.  The one-particle spectra  exhibit
a pseudogap in the total DOS and in the $K$ dependent DOS at  $(0,\pi)$ point in
BZ, similar to the hole doped spectra of the two-band Hubbard model.  The
single-band t-U Hubbard Hamiltonian is particle-hole symmetric and therefore cannot
explain the one-particle spectra in the electron-doped regime of the two-band
Hubbard model.

At the end of this section we  conclude the following:
A single-band t-U Hubbard model (i.e. $t'=0$) with a larger value of the hopping 
parameter for the hole doped regime {\em cannot} explain the 
electron-hole  asymmetries observed in the two-band Hubbard model, especially
the ones which characterize the one-particle spectral functions and the 
susceptibility functions.

\subsubsection{The t' dependence}
\label{sssec:tprim}

\begin{figure}[t]
\centerline{
\includegraphics*[width=3.3in,height=2.2in]{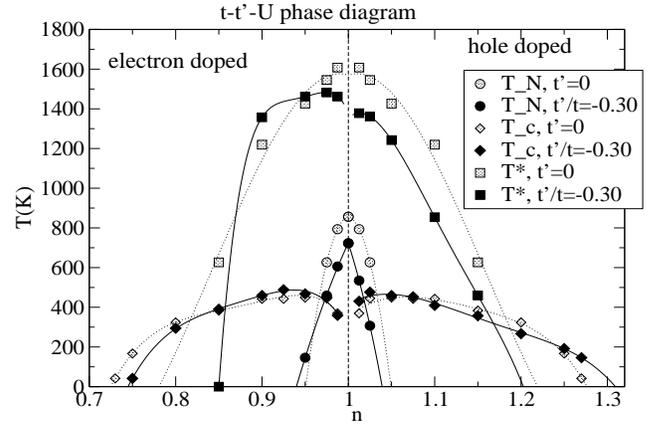}}
\caption{t-t'-U Hubbard model (solid line) and t-U Hubbard model (dashed line) phase diagrams for
$t=-0.45~eV,  U=3.6~eV $. For the t-t'-U  Hubbard model $t'/t=-0.3$. }
\label{fig:ttUphase}
\end{figure}

\begin{figure}
\centerline{
\includegraphics*[width=3.3in,height=2.2in]{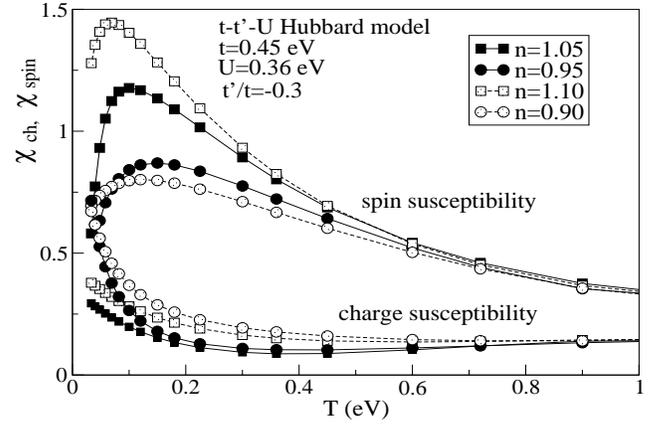}}
\caption{t-t'-U Hubbard model. Uniform spin $\chi_{spin}$  (upper part) and charge $\chi_{ch}$ (lower part)  
susceptibilities versus 
temperature for different hole densities.}
\label{fig:ttUspinsus}
\end{figure}

\begin{figure}[t]
\centerline{
\includegraphics*[width=3.3in,height=2.2in]{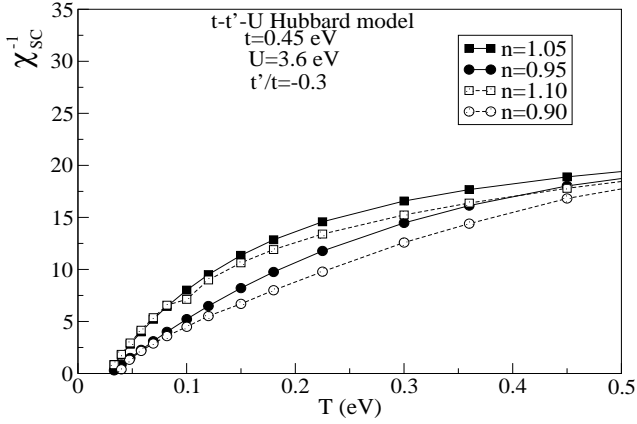}}
\caption{t-t'-U Hubbard model.
Inverse of the $d$-wave pairing  susceptibility $\chi^{-1}_{SC}$ versus temperature
for different hole densities.}
\label{fig:ttUscsus}
\end{figure}

\begin{figure}
\centerline{
\includegraphics*[width=3.3in,height=2.2in]{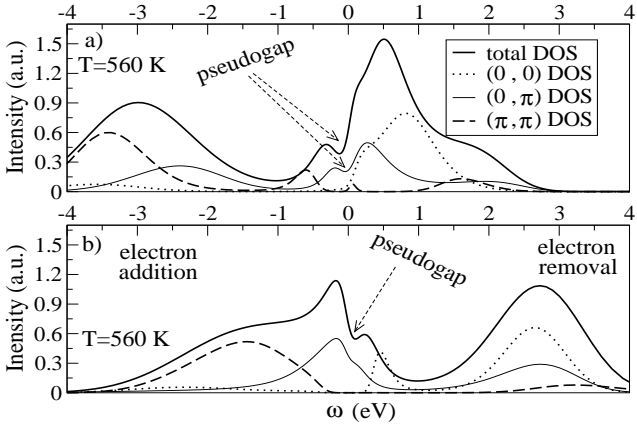}}
\caption{a) $t-t'-U$ total DOS and coarse-grained $K$ dependent DOS at $5\%$ doping for
$t=-0.45~eV, ~t'/t=-0.3, U=3.6~eV $.
a)  hole doping case. b)  electron doping case. }
\label{fig:ttUdos}
\end{figure}

In this section we study the role of the next-nearest-neighbor
hopping, $t'$, in the single-band Hubbard model
\begin{equation}
\label{eq:ttu}
H=-t \sum_{<i,j>}b^{\dagger}_ib_j-t' \sum_{<<i,j>>}b^{\dagger}_i b_j +
U\sum_{i} n_{i\uparrow} n_{i\downarrow}~.
\end{equation}

\noindent We choose the following parameters, $U=3.6~eV$,  $t=-0.45~eV$ and
$t'=0.15~eV$.  These  parameters are  close to the ones in Table.~\ref{tab:zrparam},
resulting in $J=0.22~eV$ and $J'=0.02~eV$. 

As for the two-band Hubbard model, we work in the hole representation, defined as
the one where the filling  $1+\delta$ corresponds to a hole doping $\delta$.  Values
of the filling smaller than one correspond to the electron doped regime. We keep the
sign of $t'$ always positive.  In order to avoid confusion we want to point out that
in a  t-J model the filling is always smaller than one. Therefore, in order to
describe the electron and the hole doped regimes one has to employ the hole and,
respectively, the electron representation.  Accordingly, the sign of $t'$ has to be
chosen negative in the hole doped regime and positive in the electron doped
case\footnote{Under electron-hole representation  change
both $t$ and $t'$ change sign. A change of $t$ sign has no  physical
consequences aside from a translation with $(\pi,\pi)$ in the BZ  corresponding
to a canonical transformation which changes the sign of site orbitals on one sublattice.
An equivalent statement is that the particle-hole transformation, as defined
in ref.~\cite{fradkin}, change the sign of $t'$ but not of $t$.}. 

In  Fig.~\ref{fig:ttUphase} the  phase diagram of the t-t'-U model is shown with
a solid line. For comparison, the phase diagram of t-U Hubbard model (i.e. $t'=0$
case), which is symmetric with respect to hole and electron doping, is shown with
dashed line.  At half filling $t'$ introduces an effective antiferromagnetic
exchange $J'=4t'^2/U$ between the same sublattice spins  and subsequently frustrates
the lattice.  However at finite electron doping, $t'$ favors the
antiferromagnetism, making it persist up to a larger doping.  On the other hand, in
the hole doped case, the antiferromagnetism is always suppressed by $t'$.  With
respect to superconductivity, the presence of $t'$ results in a smaller (larger)
critical electron (hole) doping at which the superconductivity disappears.  The
asymmetry introduced by $t'$ is in agreement with the one observed in the two-band
model phase diagram. We  find that $t'$ has no major influence on the maximum
superconductivity critical temperature $T_c^{max}$.

The uniform spin and charge  susceptibilities are shown in
Fig.~\ref{fig:ttUspinsus}.  The spin susceptibility at the pseudogap temperature
$T^{*}$ is strongly increasing with doping for the hole doped  case  and an opposite
effect is seen for the electron doped case.  The downturn at $T^{*}$ in the spin
susceptibility is much sharper for the hole doped regime indicating a fast
transition to the pseudogap physics. All these features are in very good qualitative agreement
with the ones corresponding to the two-band Hubbard model.  Because of the
similarity with the two-band model, it is also worth mentioning that in the
electron-doped regime the charge susceptibility is strongly increased below  $T^{*}$
in the underdoped region.

The  $d$-wave paring susceptibilities shown in  Fig.~\ref{fig:ttUscsus}  exhibit
asymmetric features, also in a qualitative agreement with those in the two-band
model.  In the electron doped regime, with increasing the doping, the pairing
correlations above $T_c$ increase. In the hole-doped regime  close  to $T_c$,
the pairing correlations do not significantly depend  on  the doping. However, 
contrary to the  two-band model behavior,
at larger temperature an increase of pairing correlations with doping is
observed. The magnitude of this increase is smaller than in the electron-doped
case and  a larger value of $t'$ (e.g. $t' \approx 0.4 t$, not shown) will reduce it further,
improving the resemblance  with the two-band model.

In Fig.~\ref{fig:ttUdos}  we present the DOS of the t-t'-U Hubbard model at $5\%$
doping.  The one-particle spectral functions resemble the corresponding two-band
Hubbard model ones.  The presence of the $t'$ parameter  is responsible for the
location of the pseudogap in the BZ.

The necessity of the $t'$  in explaining the measured ARPES line shape and the
electron-hole asymmetry was realized early on~\cite{tprime,tohyama}.  Representing
hoppings in the same sublattice, this parameter is not severely renormalized by the
AFM background, and consequently its influence turns out to be important.  Exact
diagonalization results~\cite{tohyama} of a t-t'-J model are in agreement with ours.
The $t'$ hopping process lowers the kinetic energy and moves the quasiparticle
position from $(\pi/2,\pi/2)$ to $(0,\pi)$ in the electron doped case . The
N\'{e}el-like configurations, which do not hinder this process, are stabilized. In
the hole doped case the $t'$ hopping  does not lower the kinetic energy of
quasiparticles and it is not energetically favorable, therefore leading to a
suppression of AFM at all dopings.

The main conclusion of this section is that a one-band t-t'-U Hubbard model
describes qualitatively well the physics (i.e. the phase diagram, the one-particle
spectra and the two-particle response functions) of the two-band Hubbard model,
provided a significant value of the next-nearest neighbor hopping ($t'/t \approx 0.2
-0.5$) is  considered.  However, besides all these similarities there are also some
important differences which we  are going to emphasize in the next section.

\section{Discussion}
\label{sec:disc}

In general, the deduction of an effective low-energy Hamiltonian implies two steps.
First, defining the low-energy states and second, projecting the resolvent operator,
$G(E)=(E-H)^{-1}$, on the subspace spanned by these  low-energy
states~\cite{auerbach}.  The inverse of the projected operator can be written as
$E-H_{eff}(E)$, where $H_{eff}$ is the low-energy Hamiltonian~\footnote{$H_{eff}$
can be considered independent of $E$ if the high-energy and the low-energy states
are well separated.}. This procedure is equivalent to finding  an  Hamiltonian which
produces the same one-particle, two-particle, three-particle, etc., spectral
functions on the energy range considered to be ``low-energy''.

Rigorously, in order to prove that the one-band model is the effective Hamiltonian
which describes the two-band Hubbard model low-energy physics we should compare not
only the one-particle and the two-particle spectra, but also all higher order
correlation functions.  However, we believe that the comparison of only the
one-particle and the two-particle spectral functions is compelling enough,
especially since the  experimental information is  also obtained by
measuring the response functions behavior (and in almost all cases the two-particle
operators or the one-particle ones, as in photoemission, are involved).  It is also
true that a comparison of  the dynamic susceptibilities would be required, but with
our Quantum Monte Carlo based algorithm the calculation of these quantities for the
two-band model is extremely computational resources consuming and has not been done
yet.  However partial information about the relevant excited states is contained in
the temperature behavior of the static susceptibilities.

The main conclusion of Sec.~\ref{sec:1bm} is that a t-t'-U Hubbard model describes
qualitatively well the physics of the two-band Hubbard model, but only if a
substantial next-nearest-neighbor hopping is considered. However, the calculation in
Sec.~\ref{sssec:cell} (first row of Table.~\ref{tab:zrparam}) and the more rigorous
results  by Feiner {\em et al.}~\cite{feiner},  show that in a strict ZR picture the
next-nearest-neighbor hopping is negligible. Therefore it is difficult to explain
the two-band Hubbard model physics assuming the formation of local ZR singlets.  For
hole doped systems, a significant value of $t'$ can be obtained only if the extra
holes form nonlocal bound states with the existing Cu holes, presumably something
close to the  plaquette singlets.  Of course we have no  reasons to discard other
states spread  over even more oxygen sites, which can result in a  magnitude of  the
hopping parameters different (probably not too  much) form the one obtained by
cluster calculation (second row of Table.~\ref{tab:zrparam}).  In the electron doped
systems, the doping dependent covalency shown in Fig.~\ref{fig:ndmiu}-a, clearly
indicates that the hybridization of the Cu with the  O  states at different sites is
important. A doping dependent covalency should also imply doping dependent
parameters. 

The cluster calculation which allows  the formation of non-local (plaquette)
low-energy states, unlike the cell-perturbation (or strict ZR) approach, provides a
value of the hopping parameters which captures qualitatively  the physics of the
two-band model. However, we do not believe that finding  the exact value of the
one-band Hubbard model parameters is a  relevant or even a  well addressed problem,
because the non-locality of the low-energy states implies that the two models are
not equivalent, as was pointed out by Emery {\em et al.}~\cite{ER}.  Aside from the
similarities between the two-band and the t-t'-U Hubbard models discussed in
Sec.\ref{sssec:tprim} we also find some differences. 

For example, one important difference can be observed in the $d$-wave pairing
susceptibility (Fig.~\ref{fig:ttUscsus} and Fig.~\ref{fig:scsus}). In the two-band
Hubbard model the critical exponent $\gamma$ which defines the divergence of the
susceptibility at $T_c$, is much smaller (around $\approx 0.4$ at finite
hole doping) than the one characteristic to the one-band model (around  $\approx
0.6$), indicating larger fluctuations~\footnote{The deviation of $\gamma$ from the
mean-field value $1$ is a measure of fluctuation.}. 

Both the cell-perturbation and cluster calculation provide a larger nearest-neighbor
hopping $t$ for the hole doped region. According to the analysis presented in
Sec.~\ref{sssec:tj} this should result in both  larger $T^*$ and $T_c$. However the
two-band model results does not indicate that this is the case, the respective
critical temperatures being not very different in the electron and hole doped
regimes.

Based on our comparison we can draw the following conclusions.  The one-band Hubbard
model retains much of the two-band Hubbard model physics, but a significant
next-nearest-neighbor hopping ($t'/t \approx 0.3$) should be  provided. If the
purpose of the investigation is the study of the basic physics like the SC
mechanism, the proximity of AFM, SC and  pseudogap we believe that a one-band t-t'-U
Hubbard model should be good enough.  On the other hand, if the purpose is to
describe more subtle features like the ones which may result from  the finite value
of the spin correlation on  oxygen, or if a quantitative  material specific
calculation is desired, the single-band model approach fails. Obviously also the
single-band model should not be used to describe spectral features at energies above
$0.5~eV$ such as the optical, electron energy loss and inelastic x-ray scattering
results.

\section{Summary and conclusions}
\label{sec:conc}

In this paper we use the DCA to calculate the properties of the two-band Hubbard
model.  The 2$\times$2-site cluster phase diagram resembles the generic phase
diagram of the cuprates and exhibits electron-hole asymmetry.  We also find
asymmetric features for the one-particle spectral functions and for the relevant
susceptibility functions.  These characteristics are in qualitative agreement with
experimental findings. 

We address the validity of the single-band Hamiltonian as the effective low-energy
model for the cuprates. We discuss the possible problems which may cause the failure
of the reduction from two-band to one-band and also show that, depending on the
approximations involved, the value of the one-band Hubbard parameters (especially
the next-nearest-neighbor hopping) can be significantly different.

We use DCA to study the role of the  different parameters in the single-band t-t'-U
Hubbard model and compare the phase diagram, the one-particle and the two-particle
response functions with the corresponding two-band Hubbard model ones. We conclude
that the two models exhibit similar low-energy physics provided that a significant
next-nearest-neighbor hopping $t'$ is considered. The parameter $t'$ is also the
main culprit for the electron-hole asymmetry of the cuprates.

The large value of $t'$ needed for a qualitative agreement between the two models
cannot be obtained in a strict ZR picture where the extra holes form local singlets
with the existing Cu holes. Plaquette  singlets, which in the oxygen Wannier
representation are not local, and presumably other spatially extended states can
provide a larger value of $t'$.  The doping dependent covalency in the electron
doped case also indicates that the non-local Cu-O hybridization is important.
However, the formation of non-local low-energy states also implies that they are not
real singlets and consequently cannot be rigorously mapped into holes and therefore
the two models are not equivalent. 

We also  point out some differences between the two models.  In the two-band Hubbard
model the fluctuations in the $d$-wave pairing channel above $T_c$ is much stronger.
The deduction of the parameters both in cell-perturbation and cluster approach
results in a larger nearest-neighbor hopping $t$ for the hole doped regime. However
the critical temperatures $T^*$ and $T_c$ in the two-band Hubbard model are
approximatively the same in both regimes, quite different from what should be
expected. 

The conclusion is that a single-band Hubbard model with a significant value of the
next-nearest-neighbor hopping ($t'/t \approx 0.3$), captures the basic physics of
the two-band Hubbard model, including the proximity of antiferromagnetism,
superconductivity and pseudogap and explaining the electron-hole asymmetry seen in
the phase diagram, one-particle and two-particle spectral functions. However, the
single-band Hubbard model is not entirely equivalent to the two-band Hubbard one and
we  believe that it is not suitable for quantitative material specific studies or for
describing more subtle features which may results from the non-locality of the
low-energy states.  It is also not suitable to describe physics which implies
excitations with energy scales larger than $\approx 0.5~eV$.

\section*{Acknowledgments} We thank F.C. Zhang  and Paul Kent for useful
discussions. The work was supported by NSF grant DMR-0073308, by CMSN grant DOE DE-FG02-04ER46129
and by the Netherlands
Foundation for Fundamental Research on Matter (FOM) with financial support from the
Netherlands Organization for Scientific Research (NWO) and the Spinoza Prize Program
of NWO.  The computation was performed at the Pittsburgh Supercomputer Center, the
Center for Computational Sciences at the Oak Ridge National Laboratory and the Ohio
Supercomputer Center. Part of this research was performed by TM as a Eugene P.
Wigner Fellow and staff member at the Oak Ridge National Laboratory, managed by
UT-Battelle, LLC, for the U.S. Department of Energy under Contract
DE-AC05-00OR22725.

\end{document}